\documentclass[flushrt]{aastex}
\input{psfig.tex}  

\begin{document}

\title{ METAL ABUNDANCES OF RED CLUMP STARS IN OPEN CLUSTERS: I. NGC 6819
\footnote{Based on observations made with the Italian Telescopio Nazionale
Galileo (TNG) operated on  the island of La Palma by the Centro Galileo
Galilei of the CNAA (Consorzio Nazionale per l'Astronomia e l'Astrofisica)
at the Spanish Observatorio del Roque de los Muchachos of the Instituto de
Astrofisica de Canarias} }

\author{Angela Bragaglia$^1$, Eugenio Carretta$^2$, Raffaele G. Gratton$^2$, 
Monica Tosi$^1$, Giovanni Bonanno$^3$, Pietro Bruno$^3$, Antonio Cal\`\i$^3$, 
Riccardo Claudi$^2$, Rosario Cosentino$^3$, Silvano Desidera$^4$, 
Giancarlo Farisato$^2$, Mauro Rebeschini$^2$, Salvo Scuderi$^3$}

\affil{$^1$Osservatorio Astronomico di Bologna, via Ranzani 1, 
 40127 Bologna, Italy}
\email{angela@bo.astro.it, tosi@bo.astro.it}
\affil{$^2$Osservatorio Astronomico di Padova, vicolo Osservatorio 5, 35122
Padova, Italy}
\email{carretta, gratton, claudi, farisato, rebeschini @pd.astro.it}
\affil{$^3$Osservatorio Astrofisico di Catania, Via S. Sofia 78, 95125,
Catania, Italy}
\email{gbo@sunct.ct.astro.it}
\affil{$^4$Dipartimento di Astronomia, Universit\`a di Padova, Vicolo 
Osservatorio 5, 35122 Padova, Italy}
\email{desidera@pd.astro.it}

\begin{abstract}

We present an analysis of high dispersion spectra ($R\sim 40,000$) of three red
clump stars in the old open cluster NGC 6819. The spectra were obtained with
SARG, the high dispersion spectrograph of the Telescopio Nazionale Galileo.
The spectra were analyzed using both equivalent widths measured with an
automatic procedure, and comparisons with synthetic spectra. NGC 6819 is
found to be slightly metal-rich ([Fe/H]=$+0.09\pm 0.03$, internal error);
there are no previous high resolution studies to compare with. 
Most element-to-element abundance ratios are close to solar; we find a slight
excess of Si, and a significant Na overabundance. Our spectra can also be used
to derive the interstellar reddening towards the cluster, by comparing the
observed colours with those expected from line excitation: we derive
$E(B-V)=0.14\pm 0.04$, in agreement with the most recent estimate for this
cluster.

\end{abstract}


\newpage

\section{INTRODUCTION}

Open clusters (OC's) are excellent tools to infer the evolution of 
our Galaxy from both the chemical and structural points of view, because 
they provide information on the distribution of the chemical abundances in the
disk, on the average ages at various galactic locations, on the initial 
mass function, on the interactions between thin and thick disks
(e.g., Mayor 1976, Panagia and Tosi 1981, Friel 1995, Luhman et al. 2000). 
The major advantage of using OC's rather than field stars to derive these
quantities resides in the circumstance that a) their current galactic
location is not too different from where they were born, and b) the
distances, ages and metallicities of field stars located beyond a limited 
radius from the sun are highly uncertain.

OC's probably represent the best means to understand how element
abundances change with time and with location in the Galaxy. 
Several years ago, young OC's have been suggested to show steeper negative
abundance gradients with galactocentric distance than old clusters (Mayor 1976,
Panagia \& Tosi 1981). However, later extensive studies (e.g. Friel \& Janes 
1993) have suggested that the gradient slope is roughly independent 
of the cluster age. A striking characteristics of Friel \& Janes' sample is
that at each Galactic 
radius the oldest clusters happen to be also the most metal rich ones. This
result, if confirmed, would have remarkable implications on our
understanding of the Galaxy evolution, since it is opposite to any intuitive
age-metallicity relation derivable from steady-state scenarios and suggests
the existence of short, intense phenomena which may alter significantly
local evolutions.
A caveat on these results is that they can be strongly affected by 
several uncertainties related also to the lack of homogeneity in the 
derivation of the cluster parameters (age, metallicity, distance, reddening).
As discussed e.g. by
Bragaglia et al. (2000), use of literature values can lead to a confusing
picture; for instance, ages derived with different techniques/isochrones may
not only be uncertain in absolute value, but also in ranking.

To overcome this problem, we are homogeneously analysing with great
accuracy a sample of open clusters at various galactic locations and covering
a wide range in age and metallicity. We derive the age, distance, reddening 
and approximate metallicity of these systems from deep photometry combined
with the synthetic Color-Magnitude diagrams method (see e.g. 
Sandrelli et al. 1999 and references therein), while accurate element 
abundance determination is based on high-resolution spectroscopy. The
final sample of OC's necessary to allow for significant statistics on the
abundance gradient at various epochs should be of at least 30 objects.

In this paper we present the abundances derived from high-resolution
spectroscopy of the first cluster of the sample: NGC 6819, an intermediate age
cluster (age $\simeq$ 2.3--3.1 Gyr), 
with RA(2000) = 19 41, DEC(2000) = +40 12,
l$_{\rm II}$ =73.98$^{\circ}$, b$_{\rm II}$ = $8.47^{\circ}$, and
Galactocentric distance 8.2 kpc. A recent reference for photometry and
derivation of distance and age is Rosvick \& VandenBerg (1998).

The observations have been performed with SARG, the new high dispersion 
spectrograph at the Italian Telescopio Nazionale Galileo (TNG) and this is 
the first paper where data acquired with this instrument are presented.

\section{OBSERVATIONS AND DATA REDUCTION}

\begin{figure} 
\centerline{\hbox{\psfig{figure=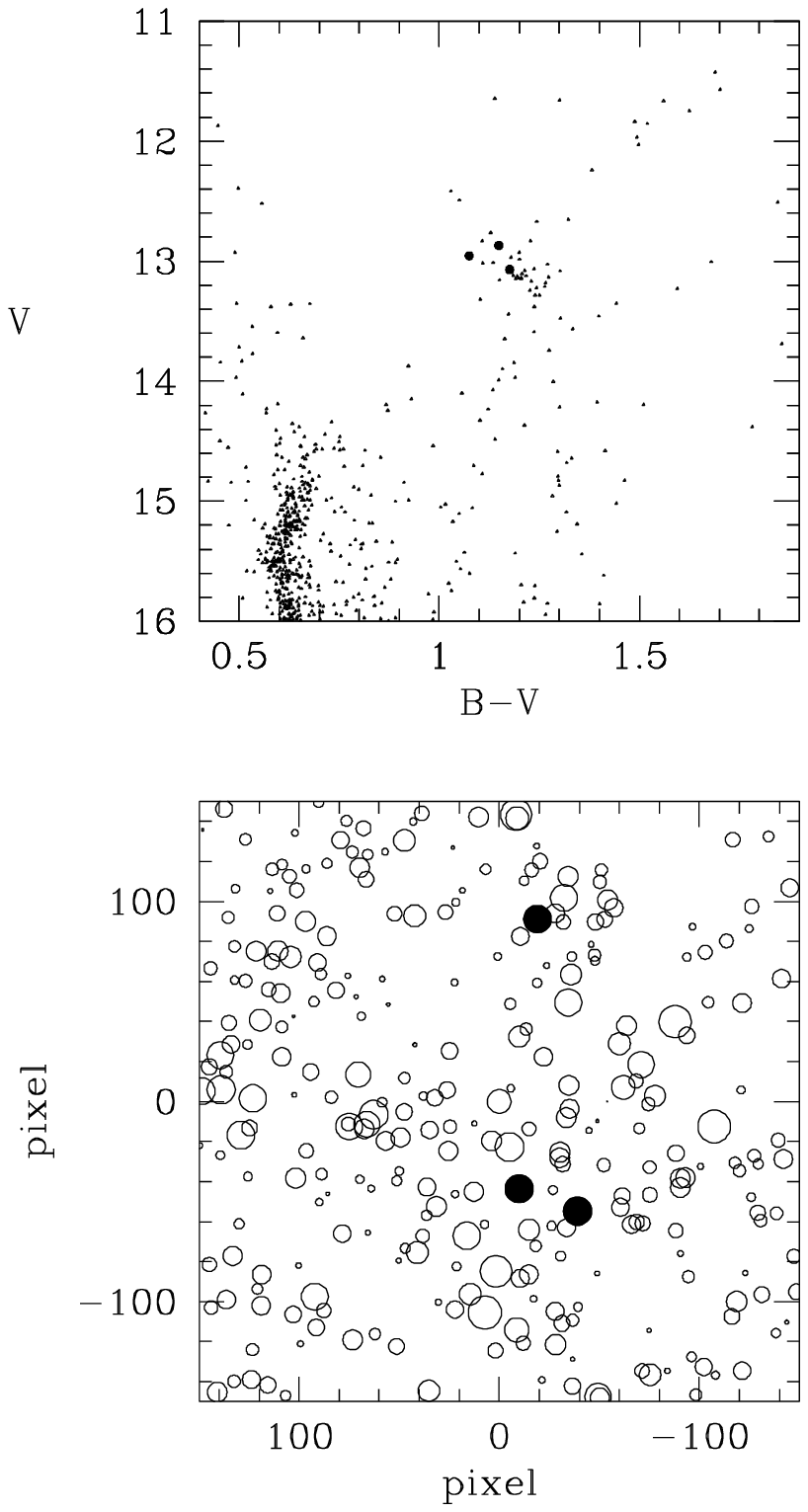,width=18.0cm,clip=}}}  
\medskip                                                                       
\caption{
Top panel: CMD for NGC 6819, taken from Rosvick \& VandenBerg
(1998). The three observed stars are shown with a heavier mark.
Bottom panel: cluster map (field of 2.8 arcmin$^2$, East on the left, 
North up) with the three observed clump stars plotted as filled circles: from
top to bottom 333, 979, and 978.
\label{f:fig1} 
}
\end{figure}

Spectra of two clump stars were obtained in July 2000, and one in August 2000,
during the second and third commissioning runs of SARG, the new high
dispersion spectrograph at the 3.5~m TNG in La Palma (Spain). The stars were
selected (using BDA, the database for Galactic open clusters, Mermilliod 1995)
among fiducial cluster members, selected from proper motions (Sanders 1972),
and now confirmed by the velocities measured from our spectra. The most
important data for the program stars are listed in Table~\ref{t:1}, along with
a summary of the spectral characteristics, and the heliocentric radial
velocities; Figure~\ref{f:fig1} gives the location of the stars in the
colour magnitude diagram (top panel) and on the sky (bottom panel).
 
Being this the first paper based on SARG data, a few details about the
spectrograph and the observations are to be provided. SARG (Gratton et al.
2000, in preparation) is a single-arm, white-pupil, cross dispersed echelle
spectrograph, permanently mounted at one of the arms of the TNG fork. It is
fed by a suitable optical train from the Nasmyth focus. The spectrograph uses
an R4 echelle grating in an off-plane, quasi-Littrow configuration, that
yields an $RS$\ product of 46,000 when coupled with the 100~mm spectrograph
beam. Spectral resolution of SARG depends on the selected slit (a set of 7 is
available on the slit wheel, from 0.27 to 1.60 arcsec), the maximum value
being about 150,000. The present observations were obtained using a slit width
of 300~$\mu$m, corresponding to 1.60~arcsec projected on the sky. However,
stellar images did not fill completely this wide slit. During the
observations, the seeing value was about 0.7 arcsec. Since the guiding
accuracy was not yet optimal (only manual guiding was available when the
program spectra were acquired), the actual images over the 1 hr exposure time
had  a FWHM of about 1.2 arcsec, yielding a spectral resolution of about
40,000. This value is confirmed by measures of the FWHM of narrow telluric
lines.
 
SARG is equipped with four grism cross dispersers mounted on a wheel; the
present observations were done using the {\it yellow} grism, which provides an
almost complete spectral coverage from 4650 to 7900~\AA, with only a small gap
of about 40~\AA ~near 6200~\AA, due to the non-sensitive region between the two
$4k\times 2k$\ thinned, back illuminated EEV CCD's located on the focal plane
of the spectrograph. Minimum interorder separation is about 7~arcsec. To
reduce read out noise and read out time, the CCD's were binned $4\times 4$.
With this choice, 2-pixel resolution is 36,000, a bit less than the FWHM of
the spectral lines, so that the spectra are slightly undersampled. The spectra
have a rather high background level, due to parasite light in the spectrograph
not yet eliminated when they were acquired, which is well removed by the
spectrum extraction procedure. The high background level has small impact on
the spectra of these rather bright stars and  the final $S/N\sim 130$\ is that
expected from photon noise.

Spectral extraction was done using standard IRAF\footnote{IRAF is distributed
by the NOAO, which are operated by AURA, under contract with NSF} routines for
echelle spectra. First, the two CCD's were separated and reduced independently.
The two dimensional images were bias subtracted (but not flat-fielded; the
flats were extracted and used to check the quality of correction done for the
blaze function, with the ISA package, Gratton 1988), then cleaned from
cosmic rays hits and hot pixels using $cosmicrays$. Images were corrected for
scattered light; the orders were then found (21 in the red CCD, and 34 in the
blue one), traced, and extracted, subtracting the local sky background.
Wavelength calibration was performed using thorium lamp exposures taken once
each night (the spectrograph is very stable, and furthermore precise radial
velocities were not our goal), yielding a $r.m.s$ of about 0.009~\AA\ in our
wavelengths.

\section{ANALYSIS AND ERROR ESTIMATES}

\subsection{Equivalent Widths}

In our analysis, we used both equivalent widths and spectral synthesis.
Equivalent widths were used to derive the best set of atmospheric parameters
(effective temperature from line excitation, gravity from the equilibrium of
ionization, microturbulence from trends with expected line strength: see
Table~\ref{t:2}) and elemental abundances. Comparisons with synthetic spectra
were then used to check the adopted abundances.

Equivalent widths (see Table~\ref{t:ew}) were measured on the unidimensional,
extracted spectral orders using an automatic routine within the ISA package,
prepared by one of us (R.G.G.). Only the spectral region 5500-7200~\AA\ was
eventually used in the abundance analysis: continuum tracing is quite
uncertain at shorter wavelengths at the resolution of our spectra, and
diffraction fringes (not completely removed by our reduction procedure) make
results from longer wavelengths uncertain. The routine measuring equivalent
widths works as follows. First, removal of the blaze function response is
performed using a spline interpolating function. Then, a local continuum level
is determined for each line by an iterative clipping average over a fraction
(the highest ones) of the 200 spectral points centered on the line to be
measured; after various tests, this fraction was set at 1/4th for the spectra
of NGC 6819. Second, the equivalent width for each line from an extensive list
(including several hundred lines, that are quite clean in the solar spectrum)
is tentatively measured using a gaussian fitting routine. Measures are
rejected according to several criteria (e.g. if the central wavelength 
does not agree with that expected from preliminary measure of 
the geocentric radial velocity; if the lines are too 
broad or too narrow, etc.). Third, the measured lines are used to
draw a relationship between equivalent width and FWHM. Fourth, this relation
is used to measure again equivalent widths using a second gaussian fitting
routine, which has only one free parameter (the equivalent width), since the
central wavelength is fixed by the measured geocentric radial velocity, and the
FWHM by the relationship between equivalent width and FWHM. Again measures are
rejected if residuals are too large, asymmetric, etc.

These procedures allow one to obtain very stable measures of the EWs, with
small random errors even when lines are in quite crowded spectral regions.
However, systematic errors may be present, related to the adopted reference
continua. Comparisons with synthetic spectra convinced us that systematic
errors in the finally adopted EWs are quite small, insofar as a suitable
fraction of the points is used to determine the reference continuum level and
rejection criteria on the used lines are kept strict.

\begin{figure} 
\centerline{\hbox{\psfig{figure=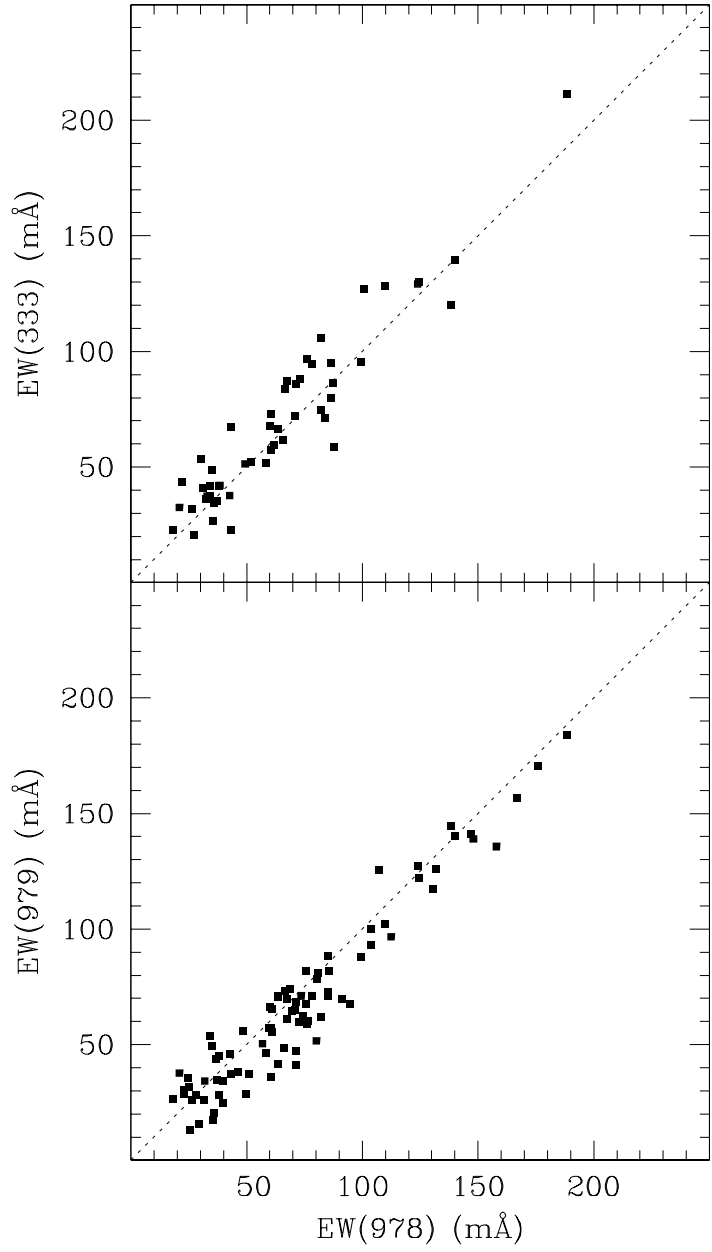,width=18.0cm,clip=}}}  
\medskip                                                                       
\caption{
Comparisons between the equivalent widths measured in star 978, and
those measured in star 333 (top panel) and 979 (bottom panel).
\label{f:fig2} 
}
\end{figure}

Since the stars have similar atmospheric parameters, we may estimate (internal)
errors in our equivalent widths by comparing values measured in different
stars. Such a comparison is shown in Figure~\ref{f:fig2}. Taking as reference
star 978 (which is the brightest, and has the best spectrum), the $rms$
scatter about the linear relationship are 10.9 and 10.5 m\AA\ for stars 333
and 979 respectively. Assuming that both sets of EWs have equal errors, we can
estimate typical errors of 7.5~m\AA ~in our measured EWs.

\subsection{Atmospheric parameters}

\begin{figure} 
\centerline{\hbox{\psfig{figure=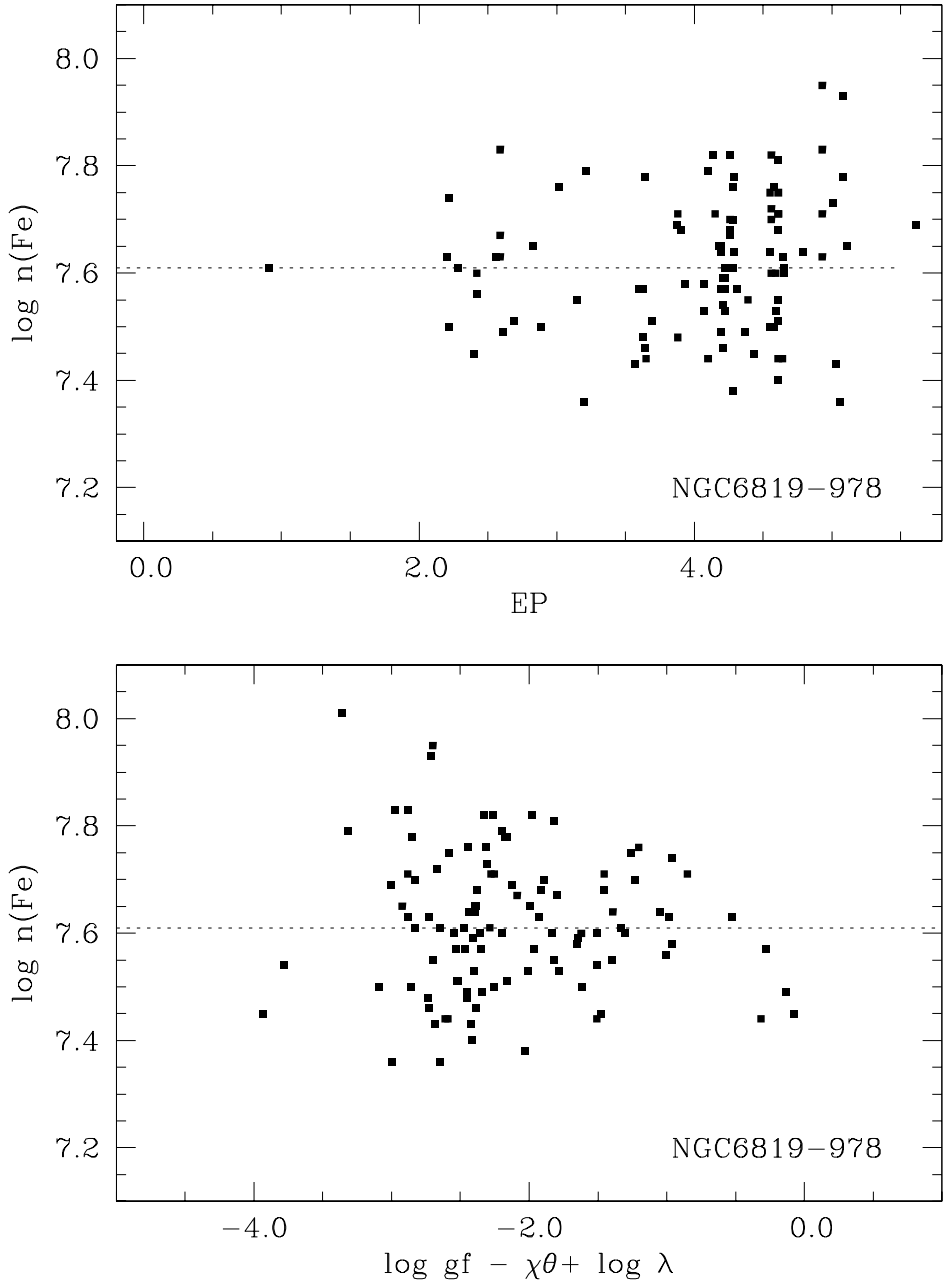,width=18.0cm,clip=}}}  
\medskip                                                                       
\caption{Trends of abundances from individual Fe I lines with excitation
potential (top panel) and expected line strength (bottom panel) for star 978.
\label{f:fig3} 
}
\end{figure}

Equivalent widths and Kurucz model atmospheres (CD-ROM 13: models with the
overshooting option switched off) were used to derive the best set of
atmospheric parameters listed in Table~\ref{t:2}. Effective temperatures were
derived directly from the spectra, imposing Fe I lines of different excitation
(several tens for each star) to provide the same temperatures. Panel (a) of
Figure~\ref{f:fig3} shows the run of abundances from Fe I lines with
excitation potential for star 978: there is no discernible trend. Internal
error estimates in temperatures are obtained directly from the errors in the
linear regression fits, and are $\pm$68 K, corresponding to 1$\sigma$ $rms$
uncertainty of 0.015 dex/eV in the slope. Systematic errors might be larger
and mainly ascribed to the set of adopted model atmospheres (Kurucz 1995).

In addition, a number of Fe II lines were measured in each star, allowing
to estimate surface gravities $\log g$\ from the equilibrium of ionization
of Fe, which is very well derived from our Fe lines. Internal errors include
contribution from both the above uncertainties in the adopted temperature
and the errors in EWs of individual lines: the quadratic sums provide an
internal error of $\pm 0.19$ dex in $\log g$. Our use of the LTE approximation
in the analysis is not of concern for these stars (see Gratton et al. 1999);
however, systematic errors can be present, mainly related again to the
adopted model atmospheres. 

The following exercise shows that the assumed temperature scale leads
to a reasonable internal consistency of data. The (average) absolute 
magnitude of the program stars is $M_V=+0.6$, according to
their apparent magnitude and to the distance modulus of $(m-M)_V=12.35$ by
Rosvick \& VandenBerg (1998), and in agreement with the average value for
clump stars measured by Hipparcos. The bolometric correction appropriate for
the temperature, gravity and metal abundances of the program stars, derived
from the Kurucz CD-ROM's is $BC=-0.44$~mag. This leads to an assumed luminosity
of the program stars of $\log L/L_\odot=1.82$, and, assuming a mass of
1.5--1.6~$M_\odot$\ (appropriate for a clump star in a 2.5~Gyr old cluster:
see Rosvick \& VandenBerg 1998), we obtain an expected gravity for the
program stars of $\log g=2.5$, in reasonable agreement with the
observed value. Note that, instead, if we decrease $T_{\rm eff}$\ by 100~K, we 
obtain a good ionization equilibrium for a gravity of $\log g\sim 2.3$, that
implies an absolute magnitude of $M_V\sim +0.3$: too bright in our opinion 
for clump stars in an old open cluster. This exercise thus suggests that
the adopted temperature scale might be overestimated but only slightly (by 
$\simeq 50$~K); such a small error would cause our [Fe/H] value to be 
overestimated by only 0.03~dex. Hereinafter, we will assume that systematic 
errors are not larger than the internal errors of $\pm 70$~K.

Microturbulent velocities $v_t$\ were obtained by compensating any trend in
the derived abundances from Fe~I lines with the expected line strength (see
Magain 1984). Results for star 978 are graphically shown in panel (b) of
Figure~\ref{f:fig3}. Due to the large number of measured Fe I lines and to the
high quality of measurements, internal error bars in the derived values of
$v_t$ are almost negligible ($\pm 0.07$ km/s, average from the 3 stars).

\subsection{Abundances}

Table~\ref{t:abu} lists the abundances obtained for the individual stars, as
well as the average values for the cluster. Sensitivities of these abundances
to changes in the atmospheric parameters are listed in Table~\ref{t:3} and
were used to estimate the internal errors in our abundances. These errors are
given in the last column of Table~\ref{t:abu} and were obtained by summing
quadratically the contribution due to each atmospheric parameter (assuming
uncertainties of 70~K in $T_{\rm eff}$, 0.19~dex in $\log g$, $0.05$~dex in
[A/H], and $0.07$~km/s in $v_t$: last Column of Table~\ref{t:3}), and that
due to the line-to-line scatter (we assumed 0.15~dex); the value was then
divided by the square root of the number of stars used to estimate abundances.

\begin{figure} 
\centerline{\hbox{\psfig{figure=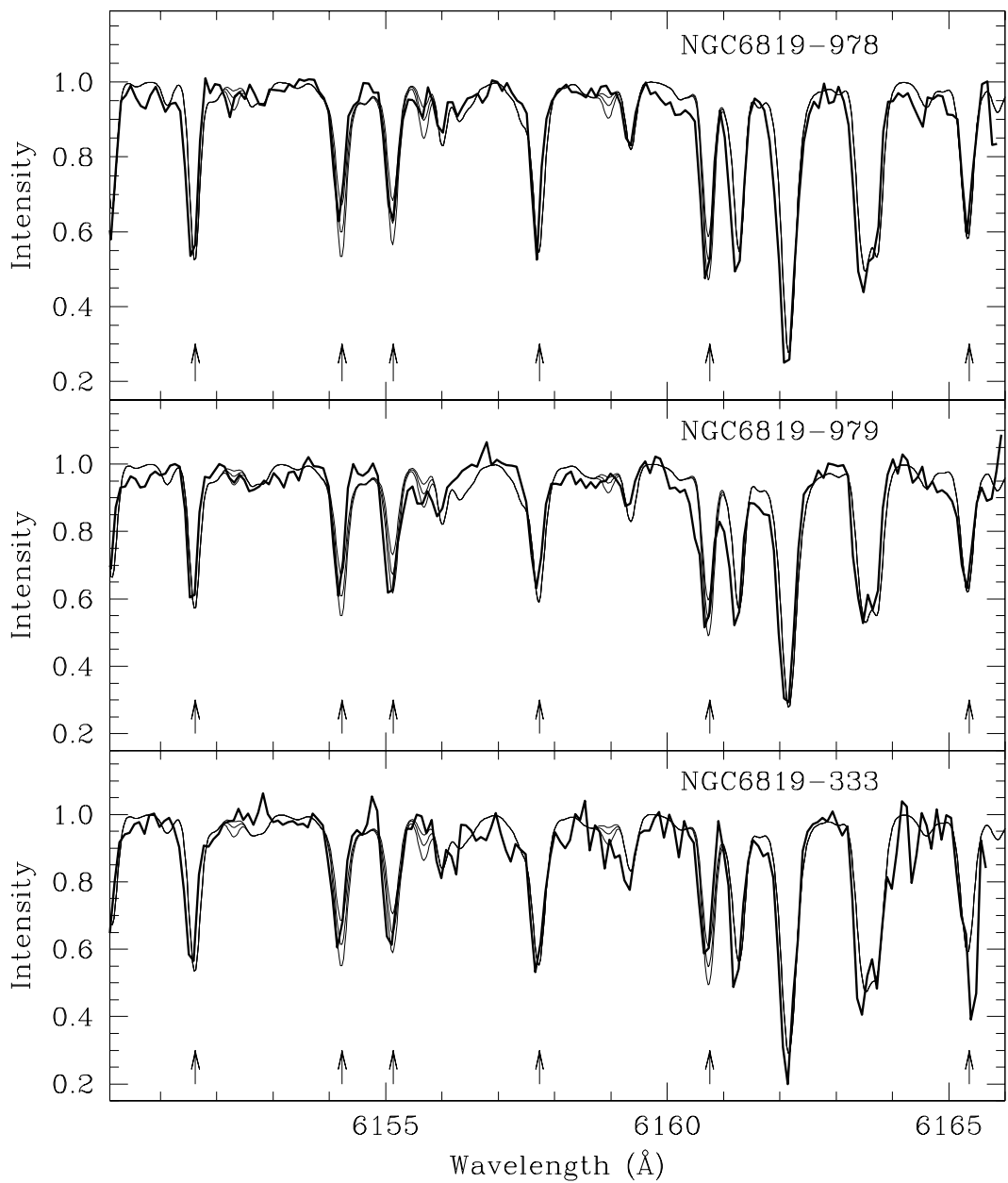,width=18.0cm,clip=}}}  
\medskip                                                                       
\caption{
Spectra of the observed stars in the region 6150-6165~\AA\
(thick lines), compared with synthetic spectra computed with the appropriate
atmospheric parameters and abundances, and three different values of Na and Si
abundances ([Na/Fe]=0.0, 0.3, and 0.6; [Si/Fe]=0.0, 0.3, and 0.6: thin lines).
Lines of Na, Si and Fe are marked (Na I 6154.230 and 6160.753, Si I 6155.135,
Fe I 6151.623, 6157.733, and 6165.363). Note the good match of the Fe I lines.
Na and Si are clearly overabundant with respect to Fe.
\label{f:na} 
}
\end{figure}

The average Fe abundance for NGC~6819 is [Fe/H]=$+0.09\pm 0.03$, the three
stars yielding very similar results. Element-to-element abundance ratios are
generally close to solar (within twice the internal error; note that
abundances for V, Mn, and Co include corrections for the quite large hyperfine
structure of their lines). There are two exceptions:
\begin{itemize}
\item We get a slight overabundance of Si (on average, [Si/Fe]=$+0.18\pm 
0.04$). This result is obtained from a rather large number of lines and
seems reasonably sound;
\item We find a quite large Na excess (on average, [Na/Fe]=$+0.47\pm 0.07$).
This value includes a small non-LTE correction ($<0.05$~dex), computed
according to Gratton et al. (1999). A partial confirmation of the results of
the equivalent width based abundance analysis comes from comparisons of the
observed spectra with synthetic spectra. An example of such a comparison is
shown in Figure~\ref{f:na}, where the spectral region 6150-6166~\AA\ is
presented, compared with results of spectral synthesis computations. Note the
good match of the Fe I lines. Na and Si lines are clearly stronger than
predicted for a solar-scaled composition, indicating an overabundance with
respect to Fe. However, the excess we would derive from this figure is about
[Na/Fe]$\sim +0.3$, a bit lower than that given by the equivalent widths.
Similar excesses of Na have been found for population I supergiants (see e.g.
Sasselov 1986); however the clump stars in NGC~6819 are much less massive than
those considered by Sasselov.
\end{itemize}

\section{SUMMARY AND CONCLUSIONS}

We can use our temperatures, derived from line excitation
(and hence a reddening free parameter) to derive the interstellar reddening
toward NGC~6819. This can be done by comparing the observed $(B-V)$\ colours
with those predicted by models (Kurucz 1995). There might be some systematic
errors in such reddening derivations, since they would depend on the accuracy
of the colours computed by model atmospheres, and on possible errors on the
$gf$'s. However, this zero-point offset (itself a function of temperature) may
be determined by observation of a comparison sample of nearby, unreddened
clump stars, analyzed following strictly the same procedure described here. We
did this for a sample of 12 bright clump stars with accurate parallaxes from
Hipparcos (Cohen et al. 2000, in preparation). This comparison sample overlaps
in temperature and metal abundance the stars of NGC~6819, so that no
extrapolation was required. Using this procedure, we derive an average
reddening for our three clump stars of $E(B-V)=0.142\pm 0.044$\ (3 stars,
r.m.s.=0.076 mag). This value agrees very well with that obtained recently 
by Rosvick \& VandenBerg ($E(B-V)=0.16$) in their accurate study based on
good quality CCD photometry. Previous determinations where contadictory,
ranging e.g. from $E(B-V)=0.12$ (Burkhead 1971) to $E(B-V)=0.28$ (Auner 1974),
and were based on more uncertain photographic photometry. 

Our determination of the metallicity for NGC~6819 ([Fe/H]=$+0.09\pm 0.03$) is
only slightly higher than the value obtained by Rosvick \& VandenBerg (1998)
from an analysis of the color--magnitude diagram ([Fe/H]=$-0.05$), and seems to
agree very well with the value given by Twarog, Ashman \& Anthony-Twarog
(1997: [Fe/H]=$+0.07\pm 0.05$), and based on a revision of the low
dispersion spectroscopic data by Friel \& Janes (1993). In fact the latter
agreement is based on a value for 
the reddening  ($E(B-V)=0.28-0.30$) much higher than ours, derived from  
quite old data and explicitly admitted to be highly uncertain. Had they
adopted a lower value for the reddening, their metallicity would have also 
been lower ([Fe/H] $\simeq$ --0.05 or --0.10, see Rosvisk \& Vandenberg
1998), but the discrepancy is fairly modest.

Our values for the metallicity and reddening of NGC6819 are close enough to the
values presented by Rosvick \& VandenBerg that no new analysis of distance
modulus and age for this cluster seems required. We must recall, however,
that the age and distances derived from isochrone fitting to the cluster
color-magnitude diagram are highly sensitive to the adopted stellar models.
Finally, we point out that our reddening and metallicity perfectly agree 
with those recently adopted by Sarajedini (1999) in his discussion of the 
dependence of the luminosity of the clump on age and metallicity.

NGC6819 hence appears to be a cluster with galactocentric distance
(8--9 kpc) similar to that of the sun, metallicity moderately higher
and age slightly younger. These features make it consistent with standard
age-metallicity relations and abundance gradients.

\acknowledgments
{We acknowledge the use of the BDA, the open clusters database maintained by
J.C. Mermilliod. This research has made use of the SIMBAD data base, operated
at CDS, Strasbourg, France. We thank the referee for thorough
comments and useful suggestions. 
Financial support has been provided by the Italian
MURST, through COFIN-1998 under the project "Stellar Evolution". Partial
support by CNAA is also acknowledged.}

\newpage

\newpage

\begin{table}
\caption{Information on the observed stars; the RV's are heliocentric, and 
derived from about one hundred FeI lines in each spectrum.
Identification is taken from the BDA (Mermilliod 1995: an extension of
the Auner 1974 numeration), photometry from 
Rosvick \& VandenBerg (1998), and membership from Sanders (1972)}
\vspace*{5mm}
{\small
\begin{tabular}{lccccccccc}
\tableline
\tableline
\\
Ident & RA     & Dec   & V &B--V   &prob.&Date obs &UT &exptime &RV \\
      & (2000) &(2000) &   &       &memb.&         &init. &sec  &kms$^{-1}$ \\
\tableline
\\
 333  &19 41 20 &+40 12 41 &13.069 &1.176 & 92 & 2000/08/17 & 00:45:41 & 3600 & 
5.31\\    
 978  &19 41 15 &+40 11 05 &12.869 &1.149 & 90 & 2000/07/18 & 01:14:35 & 3600 & 
5.96\\   
 979  &19 41 17 &+40 11 11 &12.956 &1.075 & 91 & 2000/07/18 & 02:20:19 & 3600 & 
1.44\\   
\\
\tableline
\end{tabular}
}
\label{t:1}
\end{table}

\newpage

\begin{table}
\vspace*{5cm}
\caption{Adopted atmospheric parameters and iron abundances} 
\vspace*{5mm}
\begin{tabular}{lcccc}
\tableline
\tableline
\\
star & T$_{\rm eff}$ & $\log g$ & $v_t$ & [Fe/H]\\ 
     &  (K)          & (dex)    & (km/s) &\\
\tableline
\\
333  &  4835         &  2.61    & 1.54 & 0.11 \\
978  &  4855         &  2.60    & 1.26 & 0.11 \\
979  &  4740         &  2.72    & 0.98 & 0.11 \\
\\
\tableline
\end{tabular}
\label{t:2}
\end{table}

\newpage

\begin{table}
\caption{Equivalent widths. The first column for each star is the equivalent
width (in m\AA); the second one the abundance derived from this line 
($\log n(A)$, in the scale where $\log n(H)=12$)} 
\vspace*{5mm}
{\footnotesize
\begin{tabular}{lrrrrrrrrr}
\tableline
\tableline
Element& Wavel.& E.P.& log gf&
\multicolumn{2}{c}{333}&
\multicolumn{2}{c}{978}&
\multicolumn{2}{c}{979}\\
\tableline
NaI & 5688.22 & 2.10 & -0.37 &   0.0 & 0.00 &   0.0 & 0.00 & 151.5 & 6.32 \\
NaI & 6154.23 & 2.10 & -1.57 & 109.4 & 6.77 &   0.0 & 0.00 & 105.1 & 6.81 \\
NaI & 6160.75 & 2.10 & -1.26 & 120.2 & 6.60 & 138.4 & 6.96 & 144.6 & 7.04 \\
MgI & 5528.42 & 4.34 & -0.48 &   0.0 & 0.00 &   0.0 & 0.00 & 212.1 & 7.12 \\
MgI & 5711.09 & 4.34 & -1.71 & 138.0 & 7.50 &   0.0 & 0.00 &   0.0 & 0.00 \\
MgI & 6318.71 & 5.11 & -1.97 &   0.0 & 0.00 &  71.5 & 7.65 &  41.2 & 7.17 \\
MgI & 6319.24 & 5.11 & -2.20 &   0.0 & 0.00 &   0.0 & 0.00 &  37.0 & 7.32 \\
AlI & 6696.03 & 3.14 & -1.32 &   0.0 & 0.00 &  80.2 & 6.44 &  51.6 & 5.95 \\
AlI & 6698.67 & 3.14 & -1.62 &  58.1 & 6.35 &   0.0 & 0.00 &  60.6 & 6.40 \\
SiI & 5645.62 & 4.93 & -2.14 &  62.0 & 7.72 &  65.9 & 7.83 &   0.0 & 0.00 \\
SiI & 5665.56 & 4.92 & -2.04 &   0.0 & 0.00 &   0.0 & 0.00 &  73.2 & 7.98 \\
SiI & 5684.49 & 4.95 & -1.65 &  92.8 & 7.76 &   0.0 & 0.00 &  88.8 & 7.90 \\
SiI & 5690.43 & 4.93 & -1.87 &  58.4 & 7.39 &  87.6 & 7.95 &   0.0 & 0.00 \\
SiI & 5701.11 & 4.93 & -2.05 &  67.8 & 7.72 &  60.0 & 7.63 &  66.4 & 7.87 \\
SiI & 5772.15 & 5.08 & -1.75 &   0.0 & 0.00 &  85.1 & 7.93 &  88.5 & 8.13 \\
SiI & 5793.08 & 4.93 & -2.06 &  66.7 & 7.71 &  63.8 & 7.71 &  71.2 & 7.96 \\
SiI & 5797.87 & 4.95 & -2.05 &   0.0 & 0.00 &   0.0 & 0.00 &  53.7 & 7.65 \\
SiI & 5948.55 & 5.08 & -1.23 &   0.0 & 0.00 & 107.1 & 7.78 & 125.5 & 8.16 \\
SiI & 6125.03 & 5.61 & -1.57 &  51.3 & 7.70 &  49.4 & 7.69 &   0.0 & 0.00 \\
SiI & 6145.02 & 5.61 & -1.44 &   0.0 & 0.00 &   0.0 & 0.00 &  67.4 & 8.00 \\
SiI & 6848.57 & 5.86 & -1.75 &  22.9 & 7.58 &  43.1 & 8.01 &   0.0 & 0.00 \\
S I & 6748.78 & 7.87 & -0.44 &   0.0 & 0.00 &  14.8 & 7.45 &   0.0 & 0.00 \\
S I & 6757.20 & 7.87 & -0.29 &   0.0 & 0.00 &  19.8 & 7.54 &   0.0 & 0.00 \\
CaI & 5867.57 & 2.93 & -1.49 &  65.4 & 6.35 &   0.0 & 0.00 &  49.8 & 6.11 \\
CaI & 6439.08 & 2.52 &  0.39 & 211.4 & 5.97 & 188.5 & 5.91 & 183.7 & 5.79 \\
CaI & 6449.82 & 2.52 & -0.50 & 173.1 & 6.50 &   0.0 & 0.00 &   0.0 & 0.00 \\
CaI & 6455.60 & 2.52 & -1.29 & 118.9 & 6.51 &   0.0 & 0.00 &   0.0 & 0.00 \\
CaI & 6499.65 & 2.52 & -0.82 &   0.0 & 0.00 &   0.0 & 0.00 & 117.9 & 6.21 \\
CaI & 6572.80 & 0.00 & -4.32 &   0.0 & 0.00 & 124.4 & 6.69 &   0.0 & 0.00 \\
\tableline
\end{tabular}
}
\label{t:ew}
\end{table}

\addtocounter{table}{-1}

\begin{table}
\caption{Equivalent widths (cont...)} 
\vspace*{5mm}
{\footnotesize
\begin{tabular}{lrrrrrrrrr}
\tableline
\tableline
Element& Wavel.& E.P.& log gf&
\multicolumn{2}{c}{333}&
\multicolumn{2}{c}{978}&
\multicolumn{2}{c}{979}\\
\tableline
ScI & 5671.83 & 1.45 &  0.56 &   0.0 & 0.00 &  68.2 & 3.21 &   0.0 & 0.00 \\
ScII& 5526.82 & 1.77 &  0.19 & 115.9 & 2.98 &   0.0 & 0.00 & 101.2 & 3.12 \\
ScII& 5640.99 & 1.50 & -0.86 &  88.3 & 3.18 &   0.0 & 0.00 &   0.0 & 0.00 \\
ScII& 5657.88 & 1.51 & -0.29 &   0.0 & 0.00 & 112.5 & 3.30 &  96.7 & 3.19 \\
ScII& 5684.20 & 1.51 & -0.92 &  86.5 & 3.22 &  87.2 & 3.37 &   0.0 & 0.00 \\
ScII& 6245.62 & 1.51 & -1.05 &   0.0 & 0.00 &  85.0 & 3.38 &  71.0 & 3.30 \\
ScII& 6279.74 & 1.50 & -1.16 &   0.0 & 0.00 &  73.6 & 3.26 &  71.1 & 3.40 \\
ScII& 6604.60 & 1.36 & -1.14 &   0.0 & 0.00 &  80.9 & 3.18 &  81.1 & 3.37 \\
TiI & 5490.16 & 1.46 & -0.93 &   0.0 & 0.00 &  72.7 & 4.91 &  59.8 & 4.59 \\
TiI & 5503.90 & 2.58 & -0.19 &  59.8 & 5.15 &   0.0 & 0.00 &   0.0 & 0.00 \\
TiI & 5662.16 & 2.32 & -0.11 &   0.0 & 0.00 &  77.3 & 5.19 &   0.0 & 0.00 \\
TiI & 5689.48 & 2.30 & -0.47 &  51.8 & 4.97 &  58.3 & 5.16 &  46.2 & 4.86 \\
TiI & 5978.55 & 1.87 & -0.58 &   0.0 & 0.00 &  94.6 & 5.40 &   0.0 & 0.00 \\
TiI & 6091.18 & 2.27 & -0.42 &   0.0 & 0.00 &  66.3 & 5.18 &  48.6 & 4.78 \\
TiI & 6126.22 & 1.07 & -1.42 &  95.7 & 5.07 &  99.4 & 5.32 &   0.0 & 0.00 \\
TiI & 6554.24 & 1.44 & -1.22 &   0.0 & 0.00 &  91.1 & 5.36 &  69.8 & 4.91 \\
V I & 5627.64 & 1.08 & -0.37 & 113.0 & 4.21 &   0.0 & 0.00 &   0.0 & 0.00 \\
V I & 5670.86 & 1.08 & -0.42 & 109.0 & 4.17 &   0.0 & 0.00 &   0.0 & 0.00 \\
V I & 5703.59 & 1.05 & -0.21 & 118.6 & 4.08 &   0.0 & 0.00 &   0.0 & 0.00 \\
V I & 6081.45 & 1.05 & -0.58 & 105.8 & 4.16 &  82.4 & 3.90 &  62.0 & 3.44 \\
V I & 6090.22 & 1.08 & -0.06 & 121.7 & 3.97 &   0.0 & 0.00 & 105.3 & 3.91 \\
V I & 6243.11 & 0.30 & -0.98 &   0.0 & 0.00 &   0.0 & 0.00 & 104.0 & 3.75 \\
V I & 6251.83 & 0.29 & -1.34 &   0.0 & 0.00 & 102.4 & 4.06 &   0.0 & 0.00 \\
CrI & 5702.33 & 3.45 & -0.68 &   0.0 & 0.00 &   0.0 & 0.00 &  47.3 & 5.65 \\
CrI & 5781.19 & 3.32 & -0.88 &   0.0 & 0.00 &  48.6 & 5.77 &  56.1 & 5.89 \\
CrI & 5783.07 & 3.32 & -0.40 &  79.9 & 5.75 &   0.0 & 0.00 &   0.0 & 0.00 \\
CrI & 5784.98 & 3.32 & -0.38 &  59.5 & 5.38 &  61.7 & 5.52 &   0.0 & 0.00 \\
CrI & 5787.93 & 3.32 & -0.08 &   0.0 & 0.00 &   0.0 & 0.00 &  74.0 & 5.48 \\
CrI & 6330.10 & 0.94 & -2.87 &   0.0 & 0.00 &  85.2 & 5.51 &  72.8 & 5.24 \\
CrI & 6883.07 & 3.44 & -0.42 &  84.6 & 5.91 &   0.0 & 0.00 &  66.4 & 5.71 \\
MnI & 6013.50 & 3.07 & -0.25 & 139.6 & 5.44 & 139.9 & 5.65 & 140.2 & 5.70 \\
\tableline
\end{tabular}
}
\end{table}

\addtocounter{table}{-1}

\begin{table}
\caption{Equivalent widths (cont...)} 
\vspace*{5mm}
{\footnotesize
\begin{tabular}{lrrrrrrrrr}
\tableline
\tableline
Element& Wavel.& E.P.& log gf&
\multicolumn{2}{c}{333}&
\multicolumn{2}{c}{978}&
\multicolumn{2}{c}{979}\\
\tableline
MnI & 6016.65 & 3.07 & -0.09 & 156.4 & 5.54 &   0.0 & 0.00 &   0.0 & 0.00 \\
MnI & 6021.80 & 3.08 &  0.03 & 173.5 & 5.65 &   0.0 & 0.00 &   0.0 & 0.00 \\
FeI & 5491.84 & 4.19 & -2.24 &   0.0 & 0.00 &  38.8 & 7.64 &   0.0 & 0.00 \\
FeI & 5494.47 & 4.07 & -1.96 &   0.0 & 0.00 &  54.4 & 7.53 &   0.0 & 0.00 \\
FeI & 5522.45 & 4.21 & -1.47 &   0.0 & 0.00 &  74.2 & 7.59 &  62.5 & 7.43 \\
FeI & 5539.29 & 3.64 & -2.59 &   0.0 & 0.00 &   0.0 & 0.00 &  66.9 & 7.99 \\
FeI & 5547.00 & 4.22 & -1.85 &   0.0 & 0.00 &   0.0 & 0.00 &  70.1 & 8.00 \\
FeI & 5552.69 & 4.95 & -1.69 &   0.0 & 0.00 &   0.0 & 0.00 &  12.4 & 7.24 \\
FeI & 5560.22 & 4.43 & -1.10 &  88.1 & 7.60 &  73.2 & 7.45 &   0.0 & 0.00 \\
FeI & 5568.86 & 3.63 & -2.82 &  41.8 & 7.57 &  34.4 & 7.48 &   0.0 & 0.00 \\
FeI & 5577.03 & 5.03 & -1.49 &   0.0 & 0.00 &  21.3 & 7.43 &   0.0 & 0.00 \\
FeI & 5595.05 & 5.06 & -1.69 &   0.0 & 0.00 &  12.3 & 7.36 &   0.0 & 0.00 \\
FeI & 5608.98 & 4.21 & -2.22 &  53.4 & 7.85 &  30.4 & 7.46 &   0.0 & 0.00 \\
FeI & 5609.97 & 3.64 & -3.09 &   0.0 & 0.00 &  21.0 & 7.46 &   0.0 & 0.00 \\
FeI & 5611.36 & 3.63 & -2.84 &  41.9 & 7.60 &  38.1 & 7.57 &  45.4 & 7.73 \\
FeI & 5618.64 & 4.21 & -1.34 &  94.7 & 7.72 &  78.2 & 7.54 &   0.0 & 0.00 \\
FeI & 5619.61 & 4.39 & -1.49 &   0.0 & 0.00 &  61.1 & 7.55 &  65.4 & 7.72 \\
FeI & 5635.83 & 4.26 & -1.59 &   0.0 & 0.00 &  69.6 & 7.67 &   0.0 & 0.00 \\
FeI & 5636.70 & 3.64 & -2.53 &   0.0 & 0.00 &  63.6 & 7.78 &  41.7 & 7.35 \\
FeI & 5650.00 & 5.10 & -0.80 &   0.0 & 0.00 &   0.0 & 0.00 &  54.0 & 7.56 \\
FeI & 5651.48 & 4.47 & -1.79 &  49.8 & 7.65 &   0.0 & 0.00 &  28.4 & 7.28 \\
FeI & 5652.33 & 4.26 & -1.77 &   0.0 & 0.00 &  67.7 & 7.82 &  61.0 & 7.75 \\
FeI & 5677.69 & 4.10 & -2.55 &  32.5 & 7.67 &  21.0 & 7.44 &  37.8 & 7.83 \\
FeI & 5678.39 & 3.88 & -2.88 &   0.0 & 0.00 &  19.5 & 7.48 &   0.0 & 0.00 \\
FeI & 5680.24 & 4.19 & -2.20 &  37.7 & 7.52 &  34.0 & 7.49 &  53.6 & 7.93 \\
FeI & 5701.56 & 2.56 & -2.16 &   0.0 & 0.00 &   0.0 & 0.00 & 125.9 & 7.56 \\
FeI & 5717.84 & 4.28 & -0.98 &   0.0 & 0.00 & 103.6 & 7.76 &  93.3 & 7.67 \\
FeI & 5731.77 & 4.26 & -1.10 & 102.8 & 7.67 &   0.0 & 0.00 &  80.6 & 7.50 \\
FeI & 5738.24 & 4.22 & -2.24 &  34.3 & 7.52 &  35.7 & 7.59 &  20.4 & 7.23 \\
FeI & 5741.86 & 4.26 & -1.69 &  84.0 & 7.91 &  66.7 & 7.70 &  73.2 & 7.93 \\
FeI & 5742.96 & 4.18 & -2.26 &   0.0 & 0.00 &  39.9 & 7.65 &  24.7 & 7.32 \\
\tableline
\end{tabular}
}
\end{table}

\addtocounter{table}{-1}

\begin{table}
\caption{Equivalent widths (cont...)} 
\vspace*{5mm}
{\footnotesize
\begin{tabular}{lrrrrrrrrr}
\tableline
\tableline
Element& Wavel.& E.P.& log gf&
\multicolumn{2}{c}{333}&
\multicolumn{2}{c}{978}&
\multicolumn{2}{c}{979}\\
\tableline
FeI & 5752.04 & 4.55 & -0.92 &   0.0 & 0.00 &  85.5 & 7.64 &  81.9 & 7.67 \\
FeI & 5759.26 & 4.65 & -1.98 &   0.0 & 0.00 &  25.5 & 7.60 &  13.2 & 7.21 \\
FeI & 5760.36 & 3.64 & -2.46 &   0.0 & 0.00 &   0.0 & 0.00 &  61.5 & 7.71 \\
FeI & 5778.46 & 2.59 & -3.44 &   0.0 & 0.00 &  75.5 & 7.67 &  82.0 & 7.92 \\
FeI & 5784.67 & 3.40 & -2.53 &  66.3 & 7.42 &   0.0 & 0.00 &   0.0 & 0.00 \\
FeI & 5793.92 & 4.22 & -1.62 &   0.0 & 0.00 &  63.8 & 7.53 &  70.8 & 7.76 \\
FeI & 5806.73 & 4.61 & -0.93 &  71.3 & 7.32 &  83.8 & 7.68 &   0.0 & 0.00 \\
FeI & 5811.91 & 4.14 & -2.27 &   0.0 & 0.00 &   0.0 & 0.00 &  36.3 & 7.55 \\
FeI & 5814.81 & 4.28 & -1.81 &  67.4 & 7.75 &  43.2 & 7.38 &  37.4 & 7.27 \\
FeI & 5835.11 & 4.26 & -2.18 &   0.0 & 0.00 &  40.8 & 7.68 &   0.0 & 0.00 \\
FeI & 5837.70 & 4.29 & -2.21 &  26.6 & 7.42 &  35.6 & 7.64 &   0.0 & 0.00 \\
FeI & 5849.69 & 3.69 & -2.86 &  40.8 & 7.66 &  31.1 & 7.51 &   0.0 & 0.00 \\
FeI & 5852.23 & 4.55 & -1.36 &   0.0 & 0.00 &   0.0 & 0.00 &  51.7 & 7.45 \\
FeI & 5855.09 & 4.61 & -1.56 &  46.8 & 7.51 &   0.0 & 0.00 &  33.3 & 7.31 \\
FeI & 5856.10 & 4.29 & -1.57 &  67.7 & 7.53 &   0.0 & 0.00 &  73.1 & 7.85 \\
FeI & 5858.78 & 4.22 & -2.19 &  35.3 & 7.49 &  37.3 & 7.57 &  34.8 & 7.53 \\
FeI & 5859.60 & 4.55 & -0.63 & 117.4 & 7.78 &   0.0 & 0.00 &   0.0 & 0.00 \\
FeI & 5861.11 & 4.28 & -2.26 &  36.1 & 7.65 &  32.4 & 7.61 &   0.0 & 0.00 \\
FeI & 5862.37 & 4.55 & -0.42 &   0.0 & 0.00 &   0.0 & 0.00 & 107.3 & 7.63 \\
FeI & 5879.49 & 4.61 & -1.90 &   0.0 & 0.00 &  22.8 & 7.40 &  28.8 & 7.56 \\
FeI & 5880.02 & 4.56 & -1.85 &   0.0 & 0.00 &   0.0 & 0.00 &  27.5 & 7.42 \\
FeI & 5881.28 & 4.61 & -1.76 &   0.0 & 0.00 &  43.6 & 7.71 &   0.0 & 0.00 \\
FeI & 5902.48 & 4.59 & -1.86 &  48.7 & 7.82 &  34.8 & 7.60 &  49.4 & 7.94 \\
FeI & 5905.68 & 4.65 & -0.78 &  94.8 & 7.62 &  86.4 & 7.61 &   0.0 & 0.00 \\
FeI & 5927.80 & 4.65 & -1.07 &  71.9 & 7.50 &  71.1 & 7.60 &  65.1 & 7.56 \\
FeI & 5929.68 & 4.55 & -1.16 &  87.1 & 7.75 &  67.4 & 7.50 &  69.7 & 7.63 \\
FeI & 5930.19 & 4.65 & -0.34 & 123.0 & 7.65 &   0.0 & 0.00 & 108.4 & 7.66 \\
FeI & 5933.80 & 4.64 & -2.05 &  23.1 & 7.56 &  17.9 & 7.44 &  26.6 & 7.68 \\
FeI & 5934.66 & 3.93 & -1.08 & 128.3 & 7.70 & 109.9 & 7.58 & 102.2 & 7.54 \\
FeI & 5947.53 & 4.61 & -1.95 &  22.5 & 7.41 &   0.0 & 0.00 &   0.0 & 0.00 \\
FeI & 5969.57 & 4.28 & -2.63 &   0.0 & 0.00 &  20.7 & 7.70 &   0.0 & 0.00 \\
\tableline
\end{tabular}
}
\end{table}

\addtocounter{table}{-1}

\begin{table}
\caption{Equivalent widths (cont...)} 
\vspace*{5mm}
{\footnotesize
\begin{tabular}{lrrrrrrrrr}
\tableline
\tableline
Element& Wavel.& E.P.& log gf&
\multicolumn{2}{c}{333}&
\multicolumn{2}{c}{978}&
\multicolumn{2}{c}{979}\\
\tableline
FeI & 5976.79 & 3.94 & -1.30 &   0.0 & 0.00 &   0.0 & 0.00 & 103.9 & 7.81 \\
FeI & 5984.83 & 4.73 & -0.29 & 119.3 & 7.63 &   0.0 & 0.00 & 112.0 & 7.74 \\
FeI & 6003.02 & 3.88 & -1.02 & 129.1 & 7.60 & 124.0 & 7.71 & 127.4 & 7.85 \\
FeI & 6007.97 & 4.65 & -0.76 &  80.0 & 7.34 &  86.5 & 7.60 &   0.0 & 0.00 \\
FeI & 6008.57 & 3.88 & -0.92 & 120.4 & 7.35 &   0.0 & 0.00 & 116.7 & 7.59 \\
FeI & 6015.24 & 2.22 & -4.57 &   0.0 & 0.00 &  31.6 & 7.50 &  25.9 & 7.33 \\
FeI & 6019.37 & 3.57 & -3.14 &  43.7 & 7.84 &  21.8 & 7.43 &   0.0 & 0.00 \\
FeI & 6027.06 & 4.07 & -1.20 &  96.8 & 7.42 &   0.0 & 0.00 &   0.0 & 0.00 \\
FeI & 6034.04 & 4.31 & -2.30 &  20.9 & 7.38 &  27.3 & 7.57 &   0.0 & 0.00 \\
FeI & 6054.08 & 4.37 & -2.17 &  32.1 & 7.57 &  26.5 & 7.49 &  25.9 & 7.47 \\
FeI & 6056.01 & 4.73 & -0.46 & 116.0 & 7.75 &   0.0 & 0.00 & 109.0 & 7.86 \\
FeI & 6065.49 & 2.61 & -1.49 &   0.0 & 0.00 & 175.9 & 7.49 & 170.7 & 7.44 \\
FeI & 6078.50 & 4.79 & -0.38 & 127.2 & 7.90 & 100.8 & 7.64 &   0.0 & 0.00 \\
FeI & 6079.02 & 4.65 & -0.97 &  96.7 & 7.85 &  76.2 & 7.60 &  59.1 & 7.33 \\
FeI & 6089.57 & 5.02 & -0.87 &  73.4 & 7.75 &   0.0 & 0.00 &  60.5 & 7.67 \\
FeI & 6093.65 & 4.61 & -1.32 &  85.8 & 7.95 &  71.4 & 7.81 &  47.3 & 7.38 \\
FeI & 6094.38 & 4.65 & -1.56 &   0.0 & 0.00 &   0.0 & 0.00 &  42.8 & 7.56 \\
FeI & 6096.67 & 3.98 & -1.76 &  75.8 & 7.48 &   0.0 & 0.00 &  69.5 & 7.58 \\
FeI & 6098.25 & 4.56 & -1.81 &   0.0 & 0.00 &  49.6 & 7.82 &   0.0 & 0.00 \\
FeI & 6120.26 & 0.91 & -5.77 &   0.0 & 0.00 &  59.8 & 7.61 &  57.1 & 7.53 \\
FeI & 6137.00 & 2.20 & -2.91 & 133.5 & 7.54 &   0.0 & 0.00 &   0.0 & 0.00 \\
FeI & 6151.62 & 2.18 & -3.26 & 124.6 & 7.70 &   0.0 & 0.00 &  97.4 & 7.51 \\
FeI & 6157.73 & 4.07 & -1.26 &   0.0 & 0.00 &   0.0 & 0.00 &  95.3 & 7.72 \\
FeI & 6246.33 & 3.60 & -0.73 &   0.0 & 0.00 & 158.0 & 7.57 & 135.7 & 7.34 \\
FeI & 6252.56 & 2.40 & -1.64 &   0.0 & 0.00 & 183.5 & 7.45 &   0.0 & 0.00 \\
FeI & 6270.23 & 2.86 & -2.55 &   0.0 & 0.00 &   0.0 & 0.00 & 106.0 & 7.84 \\
FeI & 6290.55 & 2.59 & -4.27 &  37.6 & 7.70 &  42.6 & 7.83 &  46.0 & 7.88 \\
FeI & 6297.80 & 2.22 & -2.70 &   0.0 & 0.00 & 142.2 & 7.74 &   0.0 & 0.00 \\
FeI & 6301.51 & 3.65 & -0.72 &   0.0 & 0.00 & 146.9 & 7.44 & 140.9 & 7.42 \\
FeI & 6311.50 & 2.83 & -3.16 &   0.0 & 0.00 &  78.4 & 7.65 &  71.1 & 7.59 \\
FeI & 6315.81 & 4.07 & -1.67 &   0.0 & 0.00 &  75.6 & 7.58 &  67.5 & 7.51 \\
\tableline
\end{tabular}
}
\end{table}

\addtocounter{table}{-1}

\begin{table}
\caption{Equivalent widths (cont...)} 
\vspace*{5mm}
{\footnotesize
\begin{tabular}{lrrrrrrrrr}
\tableline
\tableline
Element& Wavel.& E.P.& log gf&
\multicolumn{2}{c}{333}&
\multicolumn{2}{c}{978}&
\multicolumn{2}{c}{979}\\
\tableline
FeI & 6322.69 & 2.59 & -2.38 &   0.0 & 0.00 & 130.7 & 7.63 & 117.5 & 7.53 \\
FeI & 6330.85 & 4.73 & -1.22 &   0.0 & 0.00 &   0.0 & 0.00 &  62.5 & 7.72 \\
FeI & 6335.34 & 2.20 & -2.28 &   0.0 & 0.00 & 167.0 & 7.63 & 156.6 & 7.54 \\
FeI & 6392.54 & 2.28 & -3.97 &   0.0 & 0.00 &  68.4 & 7.61 &   0.0 & 0.00 \\
FeI & 6411.66 & 3.65 & -0.60 &   0.0 & 0.00 &   0.0 & 0.00 & 156.1 & 7.46 \\
FeI & 6436.41 & 4.19 & -2.31 &  42.0 & 7.67 &  38.5 & 7.65 &   0.0 & 0.00 \\
FeI & 6518.37 & 2.83 & -2.56 &   0.0 & 0.00 &   0.0 & 0.00 & 104.0 & 7.70 \\
FeI & 6533.94 & 4.56 & -1.28 &   0.0 & 0.00 &   0.0 & 0.00 &  62.5 & 7.57 \\
FeI & 6591.31 & 4.59 & -1.95 &   0.0 & 0.00 &  28.3 & 7.53 &  28.5 & 7.55 \\
FeI & 6633.76 & 4.56 & -0.81 &   0.0 & 0.00 &  97.6 & 7.70 &   0.0 & 0.00 \\
FeI & 6667.43 & 2.45 & -4.28 &   0.0 & 0.00 &   0.0 & 0.00 &  29.9 & 7.37 \\
FeI & 6667.72 & 4.58 & -2.01 &   0.0 & 0.00 &  36.9 & 7.76 &  43.7 & 7.93 \\
FeI & 6699.14 & 4.59 & -2.02 &   0.0 & 0.00 &   0.0 & 0.00 &  37.3 & 7.81 \\
FeI & 6703.58 & 2.76 & -3.00 &   0.0 & 0.00 &   0.0 & 0.00 & 100.2 & 7.94 \\
FeI & 6704.48 & 4.22 & -2.55 &   0.0 & 0.00 &  24.5 & 7.61 &  35.7 & 7.86 \\
FeI & 6725.36 & 4.10 & -2.21 &   0.0 & 0.00 &  57.8 & 7.79 &   0.0 & 0.00 \\
FeI & 6726.67 & 4.61 & -1.05 &   0.0 & 0.00 &  69.6 & 7.44 &  64.3 & 7.43 \\
FeI & 6733.15 & 4.64 & -1.44 &   0.0 & 0.00 &  56.7 & 7.63 &   0.0 & 0.00 \\
FeI & 6750.16 & 2.42 & -2.58 &   0.0 & 0.00 & 131.8 & 7.56 & 125.9 & 7.60 \\
FeI & 6753.47 & 4.56 & -2.26 &   0.0 & 0.00 &  25.1 & 7.72 &  31.9 & 7.90 \\
FeI & 6756.55 & 4.29 & -2.69 &   0.0 & 0.00 &  22.4 & 7.78 &   0.0 & 0.00 \\
FeI & 6786.86 & 4.19 & -1.90 &   0.0 & 0.00 &  56.8 & 7.57 &  50.4 & 7.49 \\
FeI & 6793.26 & 4.07 & -2.34 &   0.0 & 0.00 &   0.0 & 0.00 &  35.7 & 7.47 \\
FeI & 6796.12 & 4.14 & -2.31 &   0.0 & 0.00 &  51.3 & 7.82 &  37.3 & 7.56 \\
FeI & 6806.86 & 2.73 & -3.14 &   0.0 & 0.00 &   0.0 & 0.00 &  84.5 & 7.70 \\
FeI & 6810.27 & 4.61 & -1.00 &   0.0 & 0.00 &  86.1 & 7.71 &   0.0 & 0.00 \\
FeI & 6820.37 & 4.64 & -1.16 &   0.0 & 0.00 &   0.0 & 0.00 &  84.5 & 7.99 \\
FeI & 6843.66 & 4.55 & -0.86 &   0.0 & 0.00 &  99.3 & 7.75 &  88.1 & 7.66 \\
FeI & 6851.64 & 1.61 & -5.22 &  57.1 & 7.74 &   0.0 & 0.00 &  49.4 & 7.65 \\
FeI & 6855.72 & 4.61 & -1.71 &   0.0 & 0.00 &  38.7 & 7.51 &   0.0 & 0.00 \\
FeI & 6858.16 & 4.61 & -0.95 &   0.0 & 0.00 &  80.6 & 7.55 &  78.3 & 7.62 \\
\tableline
\end{tabular}
}
\end{table}

\addtocounter{table}{-1}

\begin{table}
\caption{Equivalent widths (cont...)} 
\vspace*{5mm}
{\footnotesize
\begin{tabular}{lrrrrrrrrr}
\tableline
\tableline
Element& Wavel.& E.P.& log gf&
\multicolumn{2}{c}{333}&
\multicolumn{2}{c}{978}&
\multicolumn{2}{c}{979}\\
\tableline
FeI & 6861.95 & 2.42 & -3.78 &   0.0 & 0.00 &  71.3 & 7.60 &  68.5 & 7.61 \\
FeI & 6862.50 & 4.56 & -1.43 &   0.0 & 0.00 &  60.8 & 7.60 &  55.5 & 7.56 \\
FeI & 6880.63 & 4.15 & -2.23 &   0.0 & 0.00 &  49.5 & 7.71 &  28.6 & 7.31 \\
FeI & 6936.50 & 4.61 & -2.14 &   0.0 & 0.00 &  29.2 & 7.75 &  15.7 & 7.38 \\
FeI & 6960.32 & 4.59 & -1.88 &   0.0 & 0.00 &   0.0 & 0.00 &  49.1 & 7.90 \\
FeI & 6971.94 & 3.02 & -3.35 &   0.0 & 0.00 &  64.6 & 7.76 &   0.0 & 0.00 \\
FeI & 7010.35 & 4.58 & -1.84 &  37.4 & 7.54 &  33.0 & 7.50 &   0.0 & 0.00 \\
FeI & 7069.54 & 2.56 & -4.20 &   0.0 & 0.00 &  39.8 & 7.63 &  34.3 & 7.49 \\
FeI & 7114.56 & 2.69 & -3.87 &   0.0 & 0.00 &  43.6 & 7.51 &   0.0 & 0.00 \\
FeI & 7118.10 & 5.01 & -1.50 &   0.0 & 0.00 &  38.1 & 7.73 &  28.4 & 7.56 \\
FeII& 5991.38 & 3.15 & -3.55 &  52.0 & 7.49 &  52.0 & 7.55 &   0.0 & 0.00 \\
FeII& 6084.10 & 3.20 & -3.80 &   0.0 & 0.00 &  31.9 & 7.36 &  34.2 & 7.65 \\
FeII& 6113.33 & 3.21 & -4.12 &   0.0 & 0.00 &  35.6 & 7.79 &  17.4 & 7.45 \\
FeII& 6149.25 & 3.89 & -2.72 &   0.0 & 0.00 &   0.0 & 0.00 &  37.1 & 7.43 \\
FeII& 6247.56 & 3.87 & -2.32 &   0.0 & 0.00 &  76.4 & 7.69 &  60.1 & 7.64 \\
FeII& 6369.46 & 2.89 & -4.21 &   0.0 & 0.00 &  34.4 & 7.50 &   0.0 & 0.00 \\
FeII& 6383.72 & 5.55 & -2.09 &  20.6 & 7.77 &   0.0 & 0.00 &   0.0 & 0.00 \\
FeII& 6416.93 & 3.89 & -2.70 &   0.0 & 0.00 &   0.0 & 0.00 &  53.2 & 7.86 \\
FeII& 6432.68 & 2.89 & -3.58 &   0.0 & 0.00 &   0.0 & 0.00 &  54.1 & 7.61 \\
FeII& 6456.39 & 3.90 & -2.10 &   0.0 & 0.00 &  84.7 & 7.68 &   0.0 & 0.00 \\
CoI & 5647.24 & 2.28 & -1.56 &   0.0 & 0.00 &  68.8 & 5.00 &  74.0 & 5.25 \\
CoI & 6455.00 & 3.63 & -0.25 &  57.5 & 4.91 &  60.5 & 5.04 &  36.2 & 4.60 \\
NiI & 5578.73 & 1.68 & -2.57 &   0.0 & 0.00 & 118.4 & 6.53 &   0.0 & 0.00 \\
NiI & 5587.87 & 1.93 & -2.39 &   0.0 & 0.00 &  98.3 & 6.17 &   0.0 & 0.00 \\
NiI & 5589.37 & 3.90 & -1.15 &  49.9 & 6.13 &   0.0 & 0.00 &   0.0 & 0.00 \\
NiI & 5593.75 & 3.90 & -0.78 &   0.0 & 0.00 &  64.3 & 6.13 &   0.0 & 0.00 \\
NiI & 5643.09 & 4.16 & -1.25 &   0.0 & 0.00 &  39.3 & 6.37 &   0.0 & 0.00 \\
NiI & 5748.36 & 1.68 & -3.25 &   0.0 & 0.00 &  94.5 & 6.61 &  67.7 & 6.15 \\
NiI & 5760.84 & 4.10 & -0.81 &   0.0 & 0.00 &  54.7 & 6.18 &   0.0 & 0.00 \\
NiI & 5805.23 & 4.17 & -0.60 &   0.0 & 0.00 &  65.4 & 6.26 &   0.0 & 0.00 \\
NiI & 5847.01 & 1.68 & -3.44 &   0.0 & 0.00 &  74.3 & 6.37 &   0.0 & 0.00 \\
\tableline
\end{tabular}
}
\end{table}

\addtocounter{table}{-1}

\begin{table}
\caption{Equivalent widths (cont...)} 
\vspace*{5mm}
{\footnotesize
\begin{tabular}{lrrrrrrrrr}
\tableline
\tableline
Element& Wavel.& E.P.& log gf&
\multicolumn{2}{c}{333}&
\multicolumn{2}{c}{978}&
\multicolumn{2}{c}{979}\\
\tableline
NiI & 5996.74 & 4.23 & -1.06 &   0.0 & 0.00 &  46.1 & 6.39 &  38.3 & 6.29 \\
NiI & 6053.69 & 4.23 & -1.07 &   0.0 & 0.00 &   0.0 & 0.00 &  43.8 & 6.43 \\
NiI & 6086.29 & 4.26 & -0.47 &  72.3 & 6.25 &   0.0 & 0.00 &   0.0 & 0.00 \\
NiI & 6108.13 & 1.68 & -2.47 & 130.0 & 6.25 & 124.6 & 6.40 & 122.2 & 6.56 \\
NiI & 6111.08 & 4.09 & -0.83 &  72.9 & 6.42 &  60.6 & 6.28 &  57.4 & 6.33 \\
NiI & 6128.98 & 1.68 & -3.39 &  74.9 & 6.15 &  82.4 & 6.43 &   0.0 & 0.00 \\
NiI & 6130.14 & 4.26 & -0.98 &   0.0 & 0.00 &  49.6 & 6.40 &   0.0 & 0.00 \\
NiI & 6322.17 & 4.15 & -1.21 &   0.0 & 0.00 &   0.0 & 0.00 &  38.6 & 6.34 \\
NiI & 6327.60 & 1.68 & -3.08 &   0.0 & 0.00 &   0.0 & 0.00 &  95.4 & 6.53 \\
NiI & 6482.81 & 1.93 & -2.78 &   0.0 & 0.00 &   0.0 & 0.00 &  87.2 & 6.35 \\
NiI & 6586.32 & 1.95 & -2.78 &   0.0 & 0.00 & 103.5 & 6.53 & 100.0 & 6.65 \\
NiI & 6598.61 & 4.23 & -0.93 &   0.0 & 0.00 &  53.3 & 6.37 &   0.0 & 0.00 \\
NiI & 6767.78 & 1.83 & -2.06 &   0.0 & 0.00 & 148.0 & 6.45 & 138.9 & 6.45 \\
NiI & 6772.32 & 3.66 & -0.96 &   0.0 & 0.00 &   0.0 & 0.00 &  90.0 & 6.63 \\
Y II& 5544.62 & 1.74 & -1.09 &   0.0 & 0.00 &  23.2 & 2.33 &  30.5 & 2.64 \\
BaII& 6496.91 & 0.60 & -0.38 &   0.0 & 0.00 &   0.0 & 0.00 & 120.6 & 2.17 \\
\tableline
\end{tabular}
}
\end{table}

\begin{table}
\caption{Average abundances. The adopted solar abundances are in the second
column (they were obtained from an inverted solar analysis, using Kurucz model
atmospheres and equivalent widths from the literature for the same set of
lines used for the stars). For each star we give the number of lines, the
average abundance (in the scale $\log n(A)$, where $\log n(H)=12$), the r.m.s.
of individual lines around this average value, and the overabundance with
respect to Fe, save for Fe for which we give the [Fe/H] value. The last
column give the average overabundances ([Fe/H] values for Fe) for NGC6819; the
error bars were obtained as explained in the text}
\vspace*{5mm}
{\footnotesize
\begin{tabular}{lrrrrrrrrrrrrrrr}
\tableline
\tableline
El.  &Sun&
\multicolumn{4}{c}{333}&
\multicolumn{4}{c}{978}&
\multicolumn{4}{c}{979}&
$<{\rm [A/Fe]}>$\\
\tableline
Fe I &7.52&49&7.62&0.16&+0.10&90&7.60&0.12&+0.08&90&7.61&0.20&+0.09&
$+0.09\pm0.03$\\
Fe II&7.52& 2&7.63&0.14&     & 6&7.60&0.15&     & 6&7.61&0.16&     &
              \\
\\
Na I &6.23& 2&6.68&0.11&+0.35& 1&6.96&    &+0.65& 3&6.72&0.37&+0.40&
$+0.47\pm0.07$\\
Mg I &7.48& 1&7.50&    &-0.08& 1&7.65&    &+0.09& 3&7.20&0.10&-0.37&
$-0.12\pm0.07$\\
Al I &6.30& 1&6.35&    &-0.05& 1&6.44&    &+0.06& 2&6.17&0.32&-0.22&
$-0.07\pm0.07$\\
Si I &7.55& 7&7.65&0.10&+0.03& 8&7.82&0.14&+0.19& 8&7.96&0.16&+0.32&
$+0.18\pm0.04$\\
Ca I &6.18& 4&6.34&0.25&+0.06& 2&6.30&0.55&+0.04& 3&6.04&0.22&-0.23&
$-0.04\pm0.06$\\
Sc I &3.10&  & .. &    & ..  & 1&3.21&    &+0.03&  & .. &    &     &
$+0.02\pm0.13$\\
Sc II&3.10& 3&3.13&0.13&-0.07& 5&3.30&0.08&+0.12& 5&3.28&0.12&+0.06&
$+0.04\pm0.04$\\
Ti I &4.94& 3&5.06&0.09&+0.02& 7&5.22&0.16&+0.20& 4&4.78&0.14&-0.25&
$-0.01\pm0.04$\\
V I  &3.93& 5&4.12&0.10&+0.09& 2&3.98&0.12&-0.03& 3&3.70&0.24&-0.32&
$-0.09\pm0.07$\\
Cr I &5.63& 3&5.68&0.27&-0.05& 3&5.60&0.14&-0.11& 5&5.59&0.25&-0.13&
$-0.10\pm0.05$\\
Mn I &5.47& 3&5.54&0.11&-0.03& 1&5.66&    &+0.11& 1&5.70&    &+0.14&
$+0.07\pm0.08$\\
Co I &4.94& 1&4.91&    &-0.13& 2&5.02&0.03& 0.00& 2&4.93&0.45&-0.10&
$-0.08\pm0.05$\\
Ni I &6.25& 5&6.24&0.11&-0.11&16&6.37&0.14&+0.04&11&6.43&0.15&+0.09&
$+0.01\pm0.02$\\
Y II &2.24&  & .. &    &  .. & 1&2.34&    &+0.02& 1&2.64&    &+0.31&
$+0.16\pm0.10$\\
Ba II&2.34&  & .. &    &  .. &  & .. &    &  .. & 1&2.16&    &-0.27&
$-0.27\pm0.15$\\
\tableline
\end{tabular}
}
\label{t:abu}
\end{table}

\begin{table}
\caption{Sensitivity of abundances to atmospheric parameters} 
\vspace*{5mm}
{\footnotesize
\begin{tabular}{lrrrrr}
\tableline
\tableline
          & $\Delta {\rm T_{eff}}$  & $\Delta$log g  & $\Delta$[A/H]&  
$\Delta$v$_t$~~~~~     & total \\
          & +100 K   & +0.3 dex  & +0.1 dex  & +0.2 km$^{-1}$ &       \\ 
\tableline
${\rm [Fe/H]I }$&  +0.059 &  +0.016 &  +0.010 &$-$0.057 & 0.048 \\
${\rm [Fe/H]II}$&$-$0.082 &  +0.159 &  +0.037 &$-$0.061 & 0.128 \\
\\
${\rm [Na/Fe]}$ &  +0.029 &$-$0.066 &$-$0.010 &$-$0.018 & 0.129 \\
${\rm [Mg/Fe]}$ &$-$0.015 &$-$0.016 &$-$0.003 &  +0.029 & 0.121 \\
${\rm [Al/Fe]}$ &  +0.013 &$-$0.032 &$-$0.013 &  +0.016 & 0.122 \\
${\rm [Si/Fe]}$ &$-$0.078 &  +0.036 &  +0.010 &  +0.016 & 0.072 \\
${\rm [Ca/Fe]}$ &  +0.082 &$-$0.072 &$-$0.016 &$-$0.046 & 0.113 \\
${\rm [Sc/Fe]}$ &  +0.070 &$-$0.037 &$-$0.009 &$-$0.043 & 0.077 \\
${\rm [Ti/Fe]}$ &  +0.076 &$-$0.024 &$-$0.018 &$-$0.028 & 0.071 \\
${\rm [V/Fe] }$ &  +0.107 &$-$0.016 &$-$0.022 &$-$0.049 & 0.114 \\
${\rm [Cr/Fe]}$ &  +0.050 &$-$0.022 &$-$0.016 &$-$0.004 & 0.079 \\
${\rm [Mn/Fe]}$ &  +0.035 &$-$0.072 &$-$0.003 &$-$0.080 & 0.133 \\
${\rm [Co/Fe]}$ &  +0.001 &  +0.031 &  +0.003 &$-$0.008 & 0.087 \\
${\rm [Ni/Fe]}$ &$-$0.013 &  +0.034 &  +0.006 &$-$0.040 & 0.040 \\
${\rm [Y/Fe] }$ &  +0.067 &$-$0.027 &  +0.001 &  +0.043 & 0.130 \\
\tableline
\end{tabular}
}
\label{t:3}
\end{table}

\end{document}